\documentclass{emulateapj}
\usepackage{apjfonts}
\usepackage{graphicx}

\usepackage{amsmath}

\shorttitle{Formation of H\,{\footnotesize I} Clouds Behind Shock Wave}
\shortauthors{T. INOUE et al.}

\begin{document}

\title{
Formation of H\,{\footnotesize I} Clouds in Shock-compressed Interstellar Medium: Physical Origin of Angular Correlation Between Filamentary Structure and Magnetic Field
}
\author{Tsuyoshi Inoue\altaffilmark{1}, and Shu-ichiro Inutsuka\altaffilmark{2}}
\altaffiltext{1}{Division of Theoretical Astronomy, National Astronomical Observatory of Japan, Osawa 2-21-1, Mitaka, Tokyo 181-0015, Japan; tsuyoshi.inoue@nao.ac.jp}
\altaffiltext{2}{Department of Physics, Graduate School of Science, Nagoya University, Furo-cho, Chikusa-ku, Nagoya 464-8602, Japan}

\begin{abstract}
Recent observations of neutral Galactic interstellar medium showed that filamentary structures of H\,{\footnotesize I} clouds are aligned with the interstellar magnetic field.
Many interesting applications are proposed based on the alignment such as measurement of magnetic field strength through the Chandrasekhar-Fermi method and removal of polarized foreground dust emissions for the detection of inflationary polarized emission in the cosmic microwave background radiation.
However, the physical origin of the alignment remains to be explained.
To understand the alignment mechanism, we examine formation of H\,{\footnotesize I} clouds triggered by shock compression of diffuse warm neutral medium using three-dimensional magnetohydrodynamic simulations with the effects of optically thin cooling and heating.
We show that the shock-compressed diffuse interstellar medium of density $n\sim 1$ cm$^{-3}$ evolves into H\,{\footnotesize I} clouds with typical density $n\sim 50$ cm$^{-3}$ via thermal instability driven by cooling, which is consistent with previous studies.
We apply a machine vision transformation developed by Clark et al.~(2014) to the resulting column density structures obtained by the simulations in order to measure angle correlation between filamentary structures of H\,{\footnotesize I} clouds and magnetic field.
We find that the orientation of H\,{\footnotesize I} filaments depends on the environmental turbulent velocity field, particularly on the strength of shear strain in the direction of the magnetic field, which is controlled by the angle between the shock propagation direction and upstream magnetic field.
When the strain along the magnetic field is weak, filamentary components of H\,{\footnotesize I} clouds basically lie perpendicular to the magnetic field.
However, the filaments have come to align with the magnetic field, if we enhance the turbulent strain along the magnetic field or if we set turbulence in the preshock medium.
\end{abstract}


\section{Introduction}
It is widely known that neutral components of the interstellar medium (ISM) have thermally bistable equilibrium states (Field et al.~1969; Wolfire et al.~1995), namely the cold neutral medium (CNM or H\,{\footnotesize I} cloud) and the warm neutral medium (WNM or diffuse ISM).
These gases have been observed through 21cm emissions and absorptions (e.g., Heiles \& Troland 2003; Dickey et al.~2003; Fukui et a.~2014).

\begin{deluxetable*}{c|cc|cccccc} \label{t1}
\tablecaption{Model Parameters}
\tablehead{Model ID & $\Theta$ $^{a}$ & $\Delta v$ [km s$^{-1}$] $^{b}$ & $\langle B \rangle_{\rm{ps}}$ [$\mu$G] $^{c}$ &
$\langle n \rangle_{\rm{RHT}}$ [cm$^{-3}$] $^{d}$ & $\langle T \rangle_{\rm{RHT}}$ [K] $^{e}$ & $\log(\langle N \rangle_{\rm{RHT}}$ [cm$^{-2}$] $)^{f}$ & $\Delta \phi_{\rm{RHT}}$ $^{g}$}
\startdata
$\Theta$22V0 & 22.5$^\circ$ & 0.0 & 7.12 & 117 & 61.4 & 19.3 & 14.0$^\circ$\\
$\Theta$45V0 & 45.0$^\circ$ & 0.0 & 8.07 & 23.2 & 225 & 18.9 & 34.3$^\circ$\\
$\Theta$67V0 & 67.5$^\circ$ & 0.0 & 8.29 & 17.3 & 298 & 18.8 & 57.9$^\circ$\\
$\Theta$90V0 & 90.0$^\circ$ & 0.0 & 8.28 & 22.8 & 223 & 18.9 & 75.5$^\circ$\\
$\Theta$22V5 & 22.5$^\circ$ & 5.0 & 5.11 & 51.7 & 114 & 19.3 & 42.5$^\circ$\\
$\Theta$45V5 & 45.0$^\circ$ & 5.0 & 6.58 & 34.1 & 173 & 19.2 & 43.8$^\circ$\\
$\Theta$67V5 & 67.5$^\circ$ & 5.0 & 7.16 & 23.2 & 256 & 19.1 & 40.7$^\circ$\\
$\Theta$90V5 & 90.0$^\circ$ & 5.0 & 6.81 & 16.8 & 374 & 19.0 & 43.3$^\circ$\\
\enddata
\tablenotetext{a}{Angle between initial magnetic field and propagation direction of shock wave ($x$-axis).}
\tablenotetext{b}{Initial velocity dispersion in the WNM.}
\tablenotetext{c}{Average magnetic field strengths in the shocked slab ($x\in [8\mbox{ pc},12\mbox{ pc}]$) at $t=3.0$ Myr.}
\tablenotetext{d}{Average density of the linear structures identified by the RHT: $\langle n \rangle_{\rm{RHT}}\equiv \int \int \int n(x,y,z)^2\,R(y,z)\,dx\,dy\,dz/\int\,\int\,\int\,n(x,y,z)\,R(y,z)\,dx\,dy\,dz$.}
\tablenotetext{e}{Average temperature of the linear structures identified by the RHT: $\langle T \rangle_{\rm{RHT}}\equiv \int \int \int T(x,y,z)\,n(x,y,z)\,R(y,z)\,dx\,dy\,dz/\int\,\int\,\int\,n(x,y,z)\,R(y,z)\,dx\,dy\,dz$.}
\tablenotetext{f}{Average column density of the linear structures: $\langle N \rangle_{\rm{RHT}}\equiv \int \int N_{\rm res}(y,z)\,R(y,z)\,dy\,dz/\int\,\int\,R(y,z)\,dy\,dz$.}
\tablenotetext{g}{Dispersion of the angular distribution of the linear structures: $\Delta \phi_{\rm RHT}\equiv (\int\,\phi^2\,R(\phi)\,d\phi/\int\,R(\phi)\,d\phi)^{1/2}$.}
\end{deluxetable*}

It has been shown by numerical studies that CNM can be formed in a shock compressed layer of the WNM:
Hennebelle \& Perault (1999; 2000) and Koyama \& Inutsuka (2000) started this line of research based on one-dimensional hydrodynamics simulations, and found that thermal instability (Field 65; Balbus 95) can grow during the condensation process to form the CNM.
Using two-dimensional simulations, Koyama \& Inutsuka (2002) found that fragmented CNMs due to the development of the thermal instability get random velocity in the shock-compressed layer, which potentially explains the origin of turbulence in molecular clouds.
Properties of such turbulent medium thanks to the thermal bistable nature of the neutral ISM were studied both analytically and numerically (Audit \& Hennebelle 2005; Inoue \& Inutsuka 2006; Hennebelle \& Audit 2007; Hennebelle et al.~2007; Inoue et al.~2007; Inoue \& Inutsuka 2008, 2009; Stone \& Zweibel 2009; van Loo et al.~2010; Gazol \& Kim 2013; Iwasaki \& Inutsuka 2012, 2014; Wareing et al. 2016).
Formation process of molecular clouds behind shock waves have also been studied by many authors (Heitsch et al.~2006; Vazquez-Semadeni et al.~2006; Hennebelle et al.~2008; Banerjee et al.~2009; Inoue \& Inutsuka 2012; Valdivia \& Hennebelle 2014; K\"ortgen \& Banerjee 2015).

Using the Australia Telescope Compact Array and the Parkes Radio Telescope, McClure-Griffiths et al.~(2006) found that the Riegel-Crutcher H\,{\footnotesize I} cloud is composed of many hairlike filaments, which are very well aligned with the ambient magnetic field.
Recently, Clark et al.~(2014) identified many slender, linear H\,{\footnotesize I} features called ``fibers" in the Galactic Arecibo L-Band Feed Array H\,{\footnotesize I} (GALFA-H\,{\footnotesize I}) Survey data.
By developing a new machine vision transformation technique named the Rolling Hough Transform (RHT), they identified the fibers and found that the H\,{\footnotesize I} fibers are oriented along the interstellar magnetic fields probed by starlight polarizations.
They also showed that angular dispersion of the fibers can be used to measure the magnetic field strength through the Chandrasekhar-Fermi method.
Based on the observed properties such as column density and linewidth, the H\,{\footnotesize I} fibers are suggested to be the CNM with density $\sim 10$ cm$^{-3}$, temperature $\sim 200$ cm$^{-3}$, and width $\sim 0.1$ pc, and are embedded in a shell of the local bubble (Bergh\"ofer \& Breitschwerdt 2002).
By using the Hessian matrix analysis of dust emission all-sky map, Planck Collaboration XXXII (2016) showed that ridges of the ISM structures usually align with the local magnetic field.
Based on the cross correlation of the H\,{\footnotesize I} fiber angles and {\it Planck} 353 GHz polarization angles, Clark et al.~(2015) discussed that the alignment of the fibers and interstellar magnetic field would provide a new tool in the search for inflationary gravitational wave B-mode polarization in the cosmic microwave background, which is currently limited by dust foreground contamination.

From the theoretical point of view, neither the origin of the H\,{\footnotesize I} fibers nor the mechanism of the alignment of the fibers and magnetic field are yet understood.
In our previous studies using two-dimensional magnetohydrodynamics (MHD) simulations, we showed that shock sweeping of the magnetized WNM create thermally unstable gas in which fragmented H\,{\footnotesize I} clouds are formed as a consequence of the thermal instability (Inoue \& Inutsuka 2008; 2009).
The physical properties of the simulated H\,{\footnotesize I} clouds such as density and temperature potentially explain the observed H\,{\footnotesize I} fibers.

However, since our previous studies were based on the two-dimensional simulations, we have not examined morphology of the clouds nor the correlation with magnetic field orientation.
In this paper, we perform three-dimensional MHD simulations of CNM formation and discuss its morphological properties using the RHT technique.
This paper is organized as follows:
In \S 2, we provide settings of the numerical simulations.
The basic results of the simulations, in particular, the evolution of shocked WNM, are shown in \S 3.1.
In \S 3.2, we apply the RHT to simulated column density structures and examine correlation between CNM fibers and magnetic field.
In \S 3.3, a physical origin of the alignment between the fibers and magnetic field is proposed.
We summarize our results and discuss their implications in \S 4.

\section{Setup of Simulations}
The basic equations we solve are the three-dimensional ideal MHD equations with interstellar cooling, heating, and thermal conduction (see, e.g., \S 2.2 of Inoue et al.~2012 for detailed equations).
We use a net cooling function in optically thin ISM provided by Koyama \& Inutsuka (2002) that is obtained by fitting various line-emission coolings (Ly $\alpha$, C\,{\footnotesize II} 158 $\mu$m, O\,{\footnotesize I} 63 $\mu$m, etc.) and photoelectric heating by polyaromatic hydrocarbons.
We impose the ideal gas equation of state with the adiabatic index 5/3, and the isotropic thermal conductivity due to the neutral atomic collisions $\kappa = 2.5 \times 10^3$ erg cm$^{-1}$ s$^{-1}$ K$^{-1}$ is used (Parker 1953).

The numerical scheme that solves the basic hydrodynamics equations is a second-order Godunov-type finite volume scheme (Sano et al.~1999; Inutsuka et al. 2015b) and the induction equations for magnetic field is solved by using a second-order consistent method of characteristics with constrained transport technique (Clarke 1996).
We use a cubic numerical domain whose side lengths are $L_{\rm box}=20$ pc with the resolution of $\Delta x = 20$ pc$/512= 3.9 \times 10^{-2}$ pc.

The initial conditions are chosen to be a thermally stable WNM with random density fluctuations whose mean density and temperature are $n = 0.5$ cm$^{-3}$ and $T = 6313$ K, respectively (corresponding sound speed is 8.27 km s$^{-1}$).
We put isotropic density fluctuations as a seed of the thermal instability that has the Kolmogorov power spectrum as $P_{\rm {1D},\rho}\propto k^{-5/3}$ in the range $1 \le k\,L_{\rm box}/2\pi \le 128$ and has an amplitude of 30 \% to the background ($\Delta \rho/\langle \rho\rangle=0.3$).
Such a Kolmogorov-type density fluctuation is known to be ubiquitous in the ISM (Armstrong et al.~1995; Chepurnov \& Lazarian 2010).
To mimic a typical turbulent diffuse ISM (e.g., Mac Low \& Klessen 2004; Federrath 2013), we put initial turbulent velocity field in some runs with the isotropic Kolmogorov power spectrum: $P_{\rm {1D},v}\propto k^{-5/3}$ in the range $1 \le k\,L_{\rm box}/2\pi \le 128$ whose velocity dispersion is $5.0$ km s$^{-1}$ (see Table 1).

Clark et al.~(2014) showed that H\,{\footnotesize I} fibers they observed seem to be embedded in the shell of the local bubble.
Bergh\"ofer \& Breitschwerdt (2002) discussed that the local bubble could be created by 20 supernovae that have occurred during the past $\sim 10$ Myr.
According to Weaver et al.~(1977) and Tomisaka et al.~(1981), the shock speed of such a bubble can be estimated as
\begin{equation}
v_{\rm sh}\simeq15\,N_{*,20}^{1/5}\,E_{51}^{-2/5}\,n_{0.5}^{-1/5}\,t_{10}^{-2/5}\,\mbox{km s}^{-1}, \label{eqSB}
\end{equation}
where $N_{20}$ is the number of supernovae normalized by 20, $E_{51}$ is the energy of each supernova normalized by $10^{51}$ erg, $n_{0.5}$ is the density of the ISM normalized by $0.5$ cm$^{-3}$, and $t_{10}$ is the age of the supershell normalized by $10$ Myr.
In order to study the evolution of the shock compressed WNM by such a bubble, we add a colliding flow velocity field of $v_{\rm coll}=15$ km s$^{-1}$ $H(L_{\rm box}/2-x)$, where $H(x)$ is the Heaviside step function.

We initially set $B=1\,\mu$G magnetic field in the $x$-$y$ plane, which is the typical field strength in the diffuse ISM (e.g., Beck 2001).
The interrelation of magnetic field and CNM morphology formed behind a shock is a key issue in this study.
So we examine various cases of the angle $\Theta$ between the initial magnetic field and propagation direction of shock wave (or $x$-axis), i.e., cases with $\Theta=22.5^\circ,\,45.0^\circ,\,67.5^\circ,$ and $90.0^\circ$.
The model parameters are summarized in Table 1.

The periodic boundary conditions are used in $y$ and $z$ directions.
In the boundary planes at $x=0$ and $L_{\rm box}$, we set continuous flows of the WNM by imposing $f(t,x=0,y,z)=f(t=0,x=L_{\rm box}-v_{\rm col} \,t,y,z)$ and $f(t,x=L_{\rm box},y,z)=f(t=0,x=v_{\rm col}\,t,y,z)$, where $f$ is any MHD variable.

In the case of the bubble with the age 10 Myr and expansion velocity given by eq.(\ref{eqSB}), more than half of the volume of the ISM is swept up by the shock wave in the last 3 Myr.
Thus we stop the simulations at $t=3.0$ Myr.

\section{Results}
\subsection{Basic Evolution of the Shocked WNM}
\begin{figure}[t]
\epsscale{1.2}
\plotone{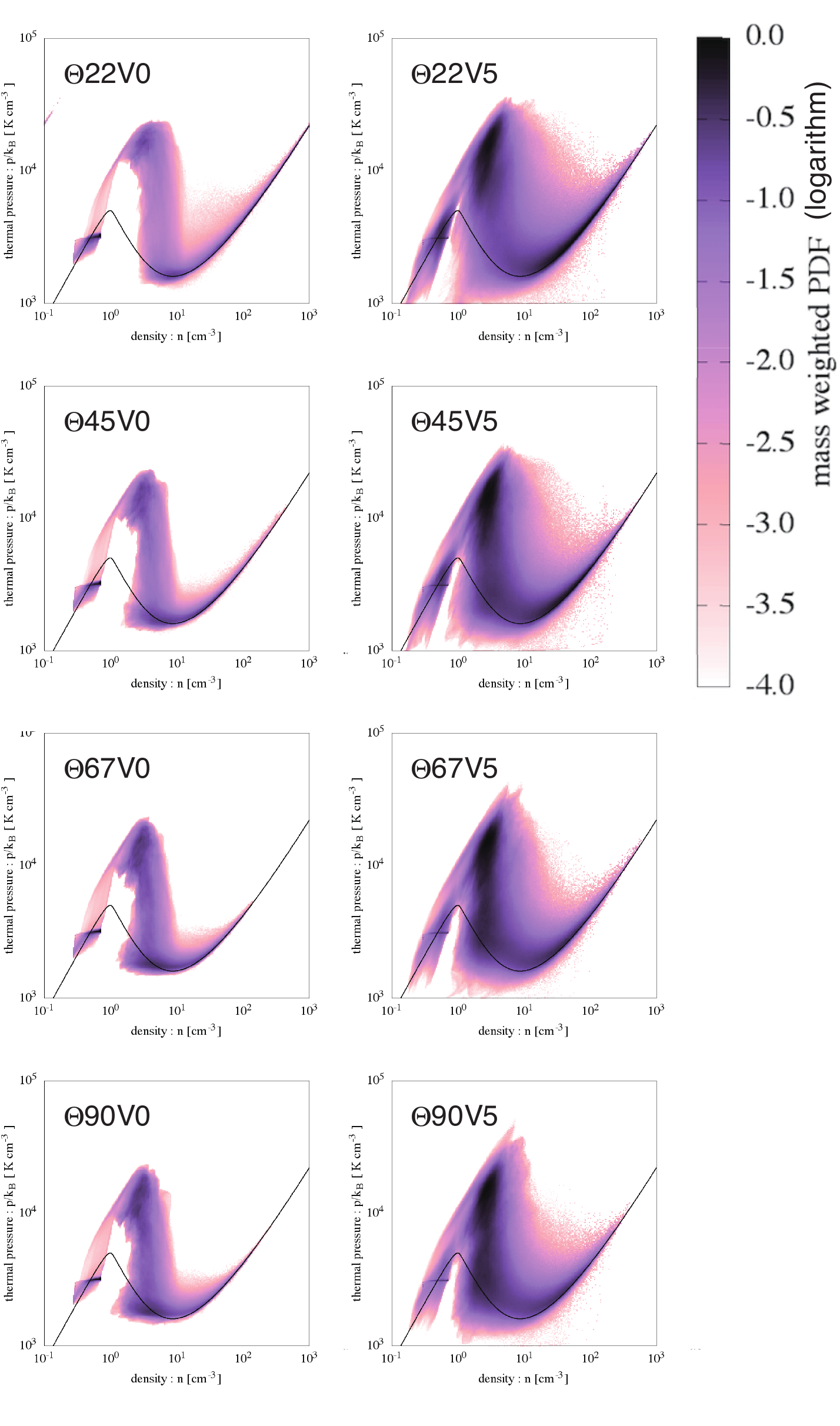}
\caption{\label{f1}
Probability distribution functions of the fluid in the number density-pressure plane at $t=3.0$ Myr.
Solid lines show the thermal equilibrium state of the cooling function.
}\end{figure}

The basic evolution and fate of the shock compressed WNM is essentially the same as that reported in our previous studies of 2D MHD simulations (Inoue \& Inutsuka 2008; 2009):
Firstly, the shock compression creates a thermally unstable gas slab in which magnetic pressure balances upstream ram pressure.
Indeed, the average magnetic field strengths in the shocked slab at $t=3.0$ Myr that are summarized in 4th row of Table 1 agree with the value expected from the pressure balance $B=(8\,\pi\,\langle \rho \rangle_{\rm wnm} v_{\rm coll}^2)=7.75\,\mu$G.
Then, the thermal instability develops to create CNM in cooling timescale of $\sim 1$ Myr.
Because the timescale of the thermal instability that enhances gas density is governed by the cooling timescale, the thermal pressure is decreased down to the initial upstream level by the time of the CMN formation.
This leads to a formation of the CNM with density $n_{\rm cnm}\lesssim 100$ cm$^{-3}$.
In Figure \ref{f1}, we show the distributions of the fluid in the number density-pressure plane at $t=3.0$ Myr.
Note that, if the initial orientation of the magnetic field is almost the same as the shock propagation direction ($\Theta=0$ case) or if we neglect magnetic field, the density of formed CNM is known to become much higher (Inoue \& Inutsuka 2008, 2009; Heitsch et al.~2009).

\begin{figure}[t]
\epsscale{1.2}
\plotone{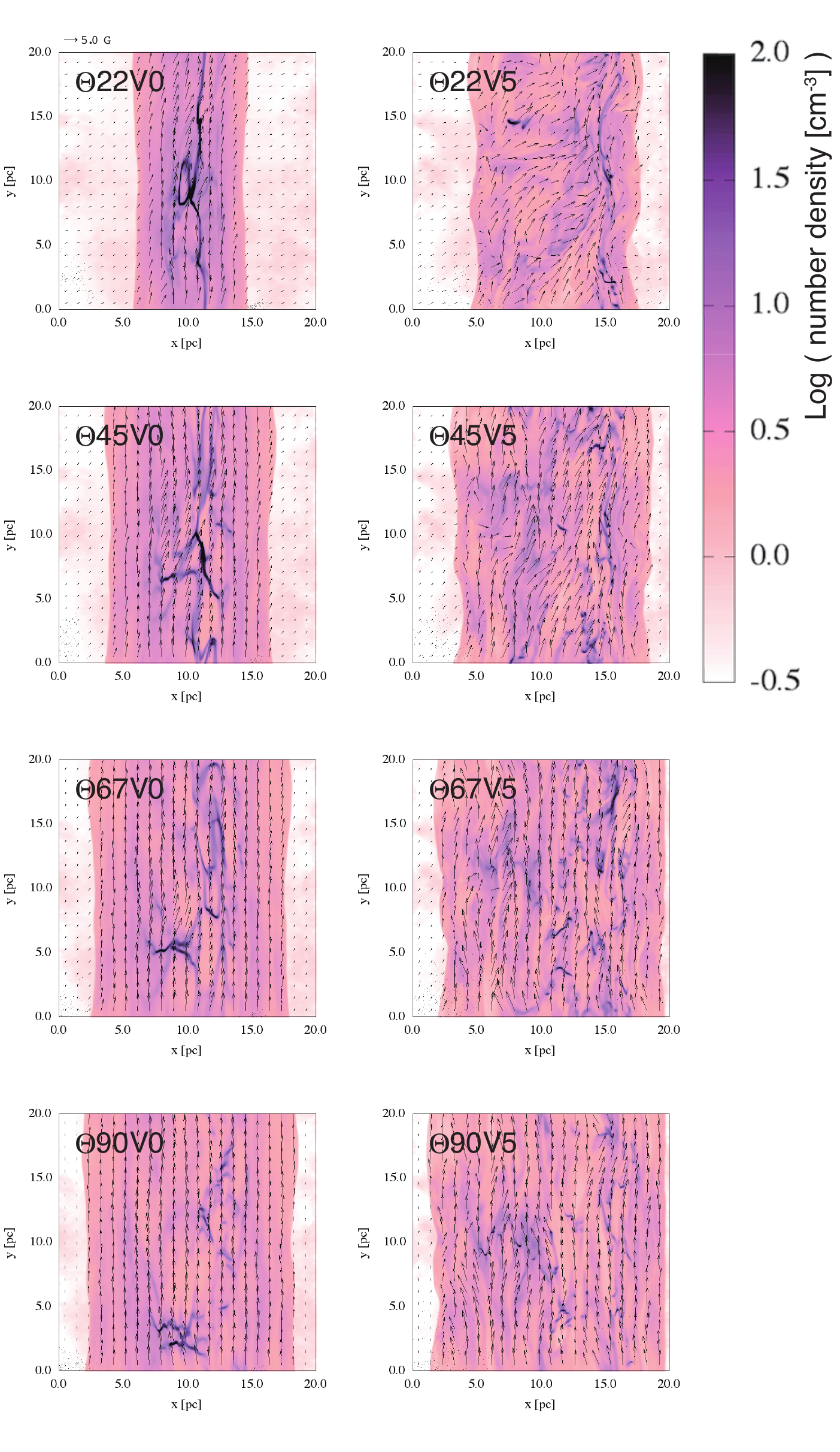}
\caption{\label{f2}
Density cross section maps at $z=0$ plane and $t=3.0$ Myr.
Arrows represent projected magnetic field vectors $(B_{x},\,B_{y})$.
}\end{figure}

In Figure \ref{f2}, we show the density cross-section maps at $z=0$ plane and $t=3.0$ Myr, in which magnetic field vectors projected onto the plane are represented as arrows.
We can see that many flattened CNM clumps are formed in the shocked slab.
The CMN are formed via the thermal instability that drives runaway gas condensation along the local magnetic field.
This is why the CNM clumps basically have a flattened shape.

\subsection{Orientation Correlation Between Filamentary CNM and Magnetic Field}
In this section, we apply the RHT developed by Clark et al.~(2014) to the column density structures of simulations and examine the orientation correlation between filamentary CNM and magnetic field.
The RHT is a method that extracts linear structures from a two-dimensional scalar map and derives their orientation.
In the RHT, there are three arbitrary parameters denoted by $D_K$, $D_W$, and $Z$.
$D_K$ is the diameter of the top hat kernel function used in the subtraction process of structures larger than the scale $D_K$.
$D_W$ is the length of the slit to detect linear structure from the data after the subtraction of the large-scale structures.
$Z$ is the coherence threshold to detect the linear structure.
Roughly speaking, $D_K$ determines the maximum width of the linear structure extracted from the column density map, $D_W$ controls the minimum length of the extracted structure, and $Z$ is the required degree of the coherency of the linear feature.
As a result of the RHT, we obtain a data cube $R(\phi,y,z)$ that has nonzero value if the linear structure is lying at the coordinate point $(y,\,z)$ with relative angle $\phi$ to the local orientation of mass weighted average magnetic field.
The exact definitions of the above parameters and a more detailed explanation of the transformation procedure are provided in Clark et al.~(2014).

Using the RHT technique, Clarke et al.~(2014) identified linear H\,{\footnotesize I} structures called ``fibers" in the GALFA-H\,{\footnotesize I} Survey data with parameters typically $D_K=10'$, $D_W=100'$, and $Z=0.7$.
If distance to the shell of the local bubble is 100 pc, their corresponding scales are $D_K\simeq0.29$ pc and $D_W=2.9$ pc.
Following these values, we adopt $D_K=0.31$ pc, $D_W=3.0$ pc, and $Z=0.7$.
We also apply the threshold column density of $10^{18}$ cm$^{-2}$ for the detection of linear features that is comparable to the sensitivity limit of the GALFA-H\,{\footnotesize I}.

\begin{figure}[t]
\epsscale{1.2}
\plotone{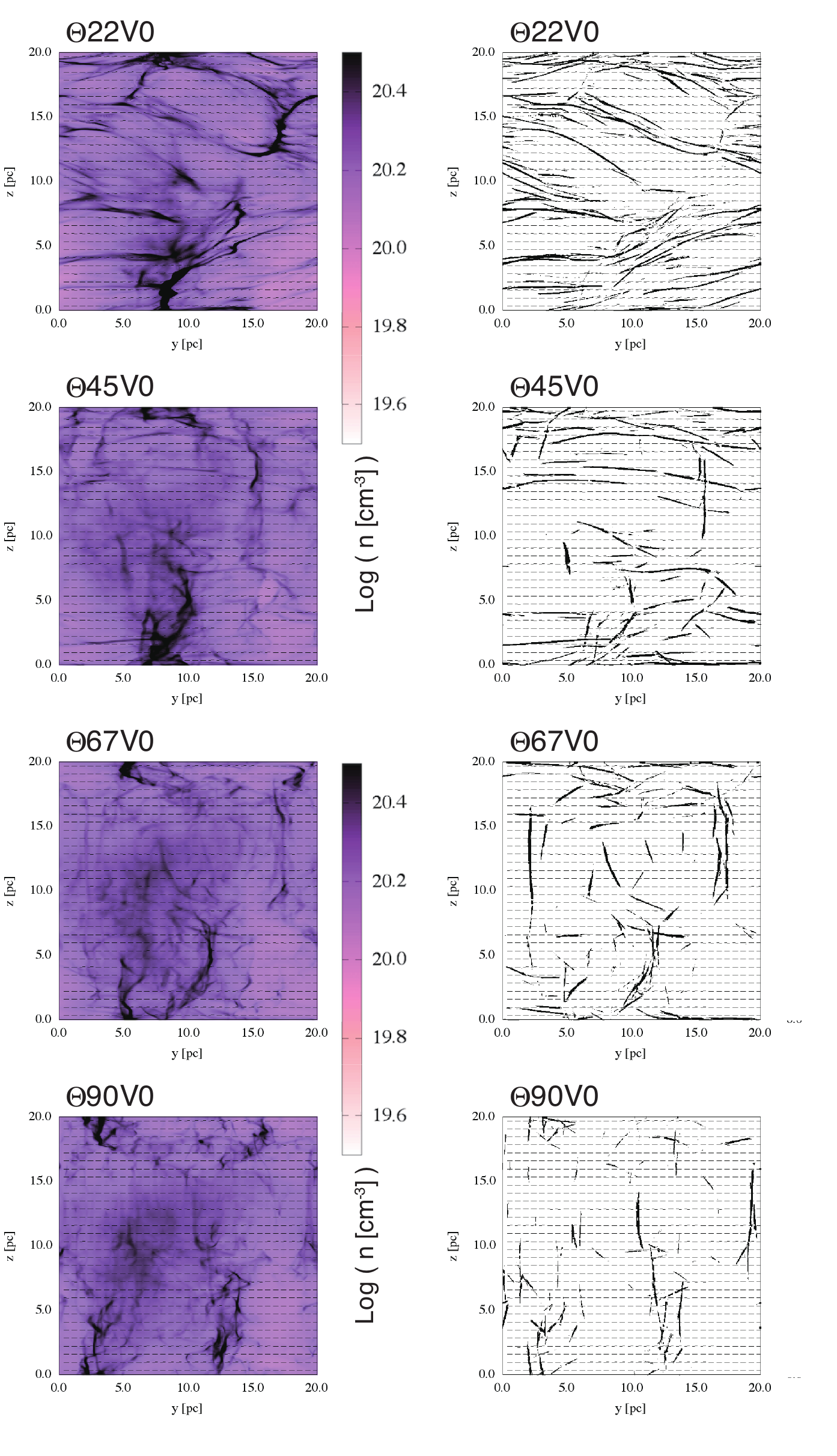}
\caption{\label{f3}
Left panels: maps of column density along the $x$-axis for the models $\Theta$22V0-$\Theta$90V0.
Right panels: back projection maps of the RHT $R(y,z)\equiv\int\,R(\phi,y,z,)\,d\phi$: grids are blacked out if the RHT detects linear structure there.
Mass weighted average magnetic fields $\langle \vec{B} \rangle =(\int\,\rho\,B_{y}\,dx/\int\,\rho\,dx,\,\int\,\rho\,B_{z}\,dx/\int\,\rho\,dx)$ are plotted as bars.
}\end{figure}

\begin{figure}[t]
\epsscale{1.2}
\plotone{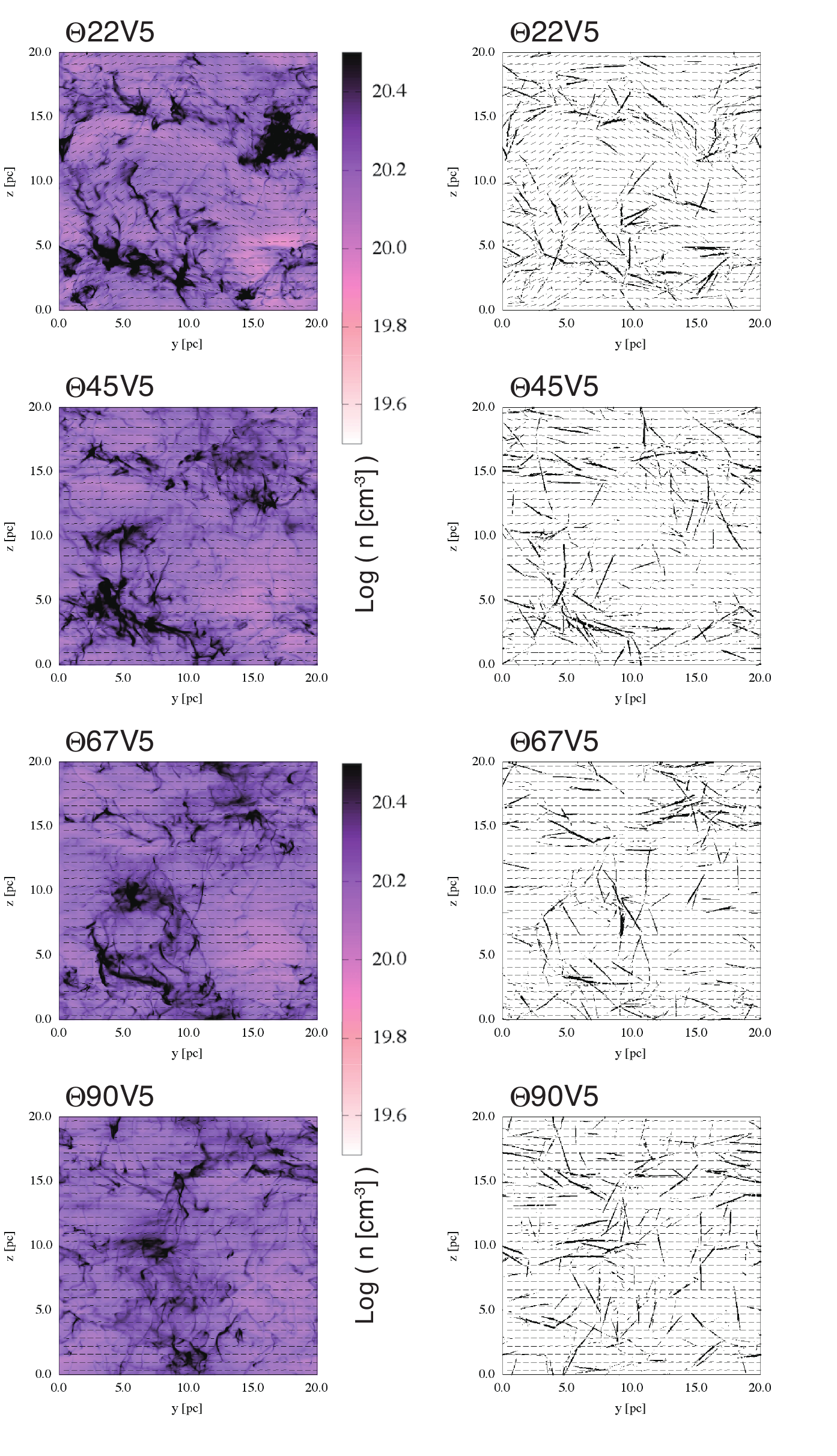}
\caption{\label{f4}
Same as Figure \ref{f3} but for the models $\Theta$22V5-$\Theta$90V5.
}\end{figure}

In left panels of Figure \ref{f3} and Figure \ref{f4}, we show the column density structures of all the runs at $t=3.0$ Myr, where projected orientations of mass weighted average magnetic fields $\langle \vec{B} \rangle =(\int\,\rho\,B_{y}\,dx/\int\,\rho\,dx,\,\int\,\rho\,B_{z}\,dx/\int\,\rho\,dx)$ are plotted as bars.
Right panels of Figure \ref{f3} and Figure \ref{f4} show back projections of the RHT $R(y,z)\equiv\int\,R(\phi,y,z,)\,d\phi$, i.e., grids are blacked if the RHT detects linear structure there.
In these Figures, the $x$-direction is chosen as the line-of-sight.
This line-of-sight selection is reasonable, because the sun is sitting roughly in the centre of the local bubble (Sfeir et al.~1999).

\begin{figure}[t]
\epsscale{1.}
\plotone{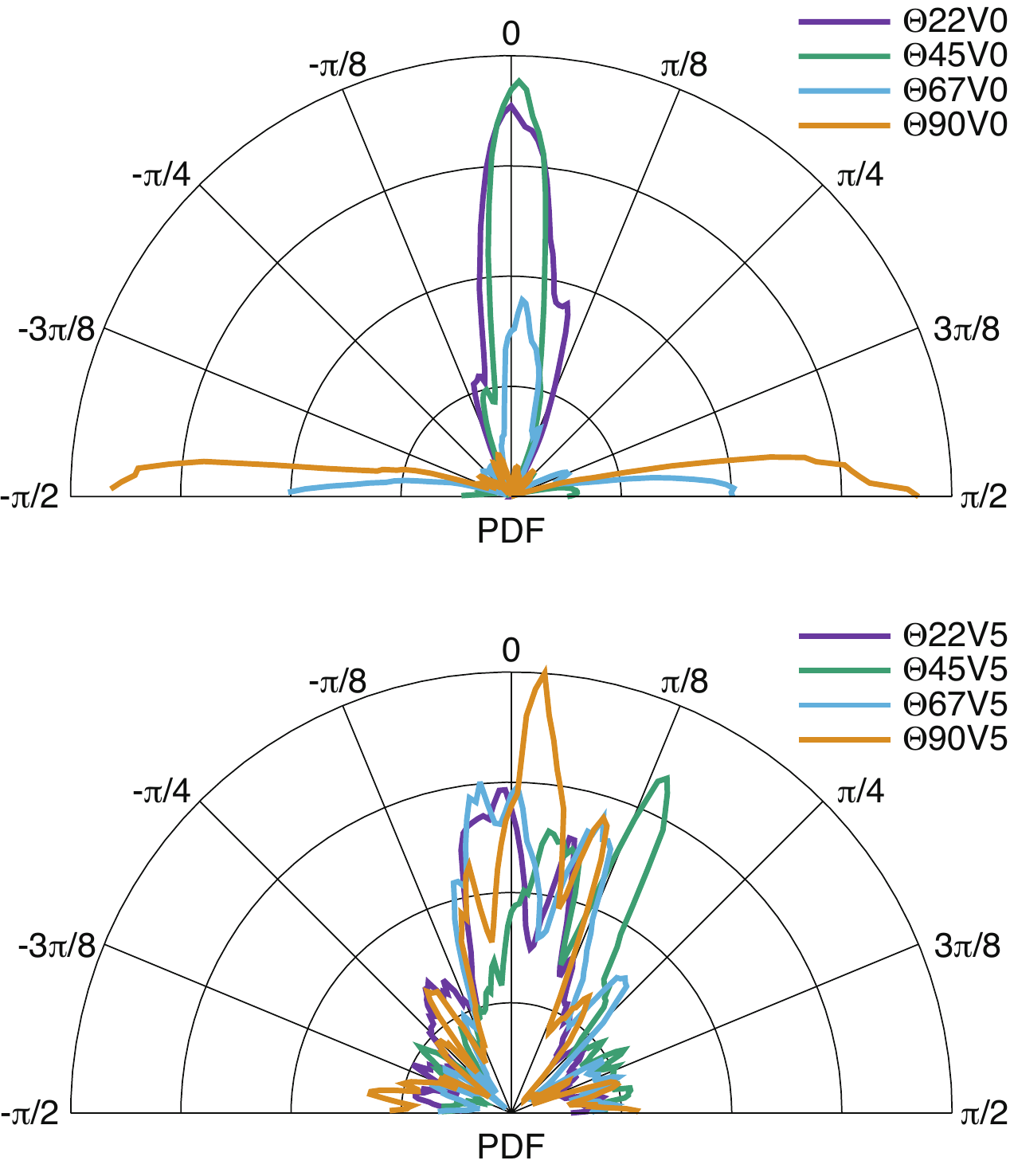}
\caption{\label{f5}
Distribution function of the angles $\phi$ between the fibers and the mass weighted average magnetic fields ($R(\phi)\equiv \int \int R(\phi,y,z)\,dy\,dz$).
Top panel shows the results of the models $\Theta$22V0-$\Theta$90V0.
Bottom panel shows the results of the models $\Theta$22V5-$\Theta$90V5.
Dispersions of $R(\phi)$ are listed in Table 1.
}\end{figure}

In Figure \ref{f3}, one may find that most fibers seem to be oriented along the magnetic field in the result of the run $\Theta22V0$, and the number of fibers that orient perpendicular to the magnetic field seems to increase with $\Theta$.
In the results of models with initial turbulence, although structures are much more complicated, Figure \ref{f4} shows that the fibers are roughly oriented in the direction of the local magnetic field.
To show these features more explicitly, in Figure \ref{f5}, we plot the angular distribution function of the fibers and the mass weighted average magnetic fields ($R(\phi)\equiv \int \int R(\phi,y,z)\,dy\,dz$).
The top panel shows the results of the models $\Theta$22V0-$\Theta$90V0, and the bottom panel shows the results of the models $\Theta$22V5-$\Theta$90V5.
The dispersion of the $R(\phi)$ is listed in Table 1.
We can confirm the above-stated features.

The mass-weighted average volume density/temperature, and average surface density of the fibers are listed in Table 1.
These values are very close to those reported in Clarke et al.~(2014).
Note that the RHT identify the linear structures from the residual map $N_{\rm res}$ that is computed by subtracting smoothed column density from original column density, i.e., $N_{\rm res}=N(y,z)-\int \int N(y',z')\,H\{D_K^2-(y-y')^2+(z-z')^2\}\,dy'\,dz'$ where $H$ is the Heaviside step function.
Thus, the column densities of fibers in Table 1 do not include the column density of large scale structures such as foreground/background diffuse component and clouds of scale larger than $D_K$.

\subsection{Mechanism of Alignment}
Why do the different angular distributions appear in models $\Theta22V0$-$\Theta90V0$?
Understanding of the difference would be directly linked to the understanding of the observed alignment of H\,{\footnotesize I} fibers and interstellar magnetic field.
In the following, we propose that differing characters of turbulence between models provide the reason.

\begin{figure}[t]
\epsscale{1.}
\plotone{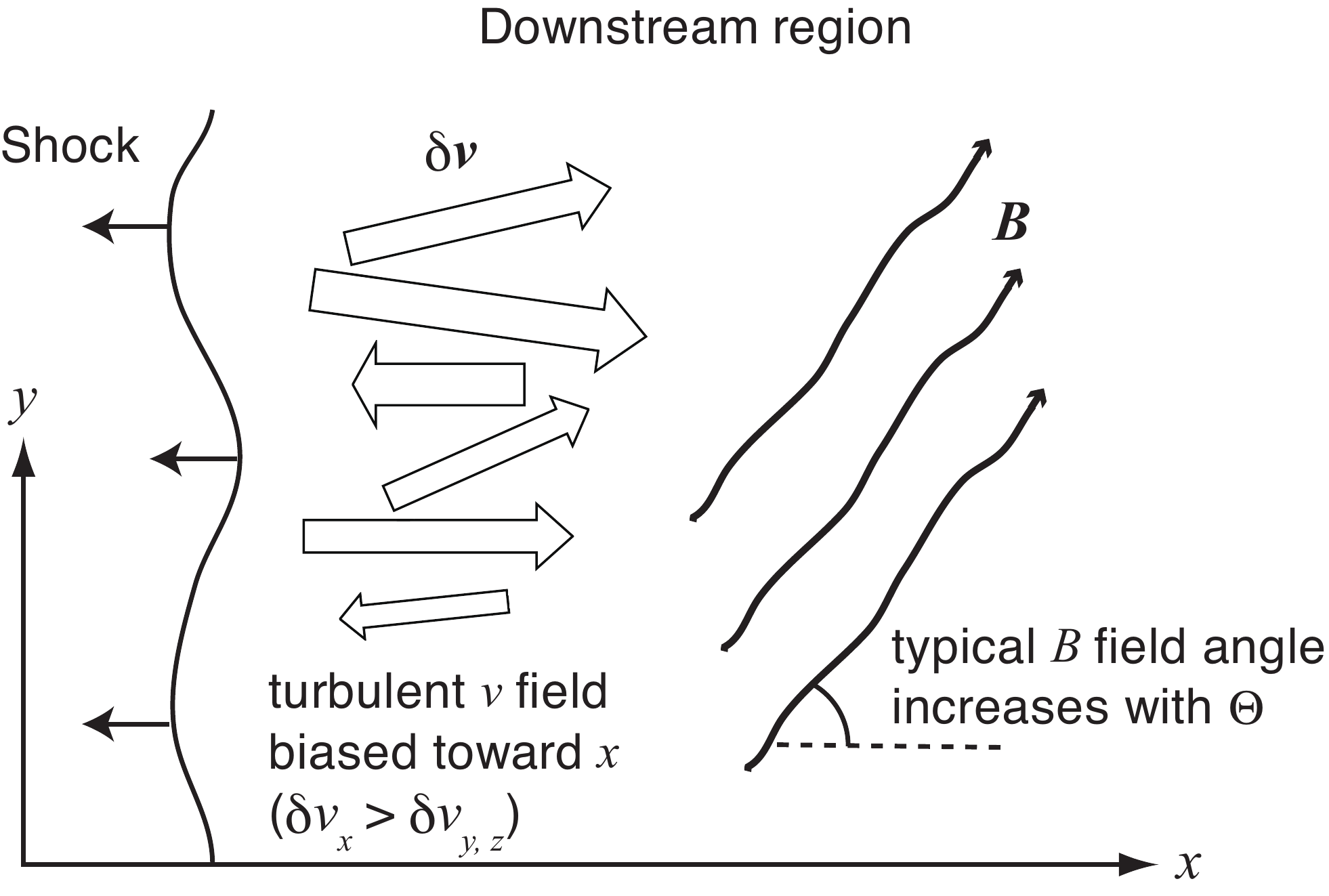}
\caption{\label{f6}
Schematic illustration of downstream region for the models with no initial turbulence ($\Theta$22V0-$\Theta$90V0).
In these models, downstream turbulence is created at shock front and has anisotropic velocity dispersion biased in the $x$-direction ($\Delta v_x>\Delta v_{y,z}$).
The typical orientation of the magnetic field behind shock is basically controlled by the initial $\Theta$.
Thus, the stretching rate of clouds along magnetic field ($\propto\partial v_x/\partial x \cos\Theta$) is expected to decrease with $\Theta$.
}\end{figure}

It is known that when gas condensation triggered by the thermal instability from a static thermally unstable medium, the CNM is formed that is flattened perpendicular to the magnetic field (see Figure 2 of Inoue et al.~2009), because magnetic pressure prevents the gas condensation except in the direction along the local magnetic field.
According to the theoretical study of shock-induced turbulence by Inoue et al.~(2013), the interaction between the shock wave and preshock density inhomogeneity drives turbulence at the shock front and velocity dispersion becomes anisotropic biased in the direction of the shock propagation, i.e., $\Delta v_x>\Delta v_{y,z}$ (see also, V\'azquez-Semadeni et al. 2007; Inoue \& Inutsuka 2012).
In the models of $\Theta$22V0--$\Theta$90V0, the postshock velocity field is dominated by the turbulence induced by the shock-density fluctuation interaction, since there is no initial turbulent velocity in these runs.
This indicates that the post shock turbulent field tends to strain post shock gas structure in the $x$-direction.
The stretching rate can be characterized by $\langle \partial v_{x}/\partial y \rangle$.
However, such a turbulent shear strain is restricted by the magnetic field, and shear motion perpendicular to the magnetic field is suppressed by magnetic tension force, leading to the modified stretching rate:
\begin{equation}
\langle \partial\,v_{B\parallel}/\partial l_{B\bot}\rangle \simeq \langle \partial v_{x}/\partial y \rangle\,\cos \Theta. \label{eqSR}
\end{equation}
The situation is illustrated schematically in Figure \ref{f6}.
Eq. (\ref{eqSR}) states that the strength of the shear strain decreases as the initial angle $\Theta$ increases.
This explains why angular distribution of the fibers changes with the initial magnetic field angle $\Theta$.

In the series of runs with the initial upstream turbulence (runs $\Theta$22V5--$\Theta$90V5), all the PDF have peaks at $\phi\sim 0$ although dispersion is large.
This is quite reasonable, because the initial turbulence is given as isotropic turbulence, and thus there is always strain force along magnetic field from the beginning.
Hennebelle (2013) showed that a similar turbulent shear strain can be the origin of non-self-gravitating filaments in the ISM.
Our results indicate that this mechanism is operating in the formation process of HI fibers.
Similar discussion was made by Planck collaboration XXXII (2016) that the turbulent shear flows could be the origin of the alignment between the magnetic fields and the ridges of the ISM structures.
They showed that the dispersion of the relative angle between the ridge structures and the local magnetic field is $33^\circ$ that is compatible with the results of the simulations starting from the moderate initial upstream turbulence (see, the dispersion of the $R(\phi)$ listed in Table 1).

\begin{figure}[t]
\epsscale{1.}
\plotone{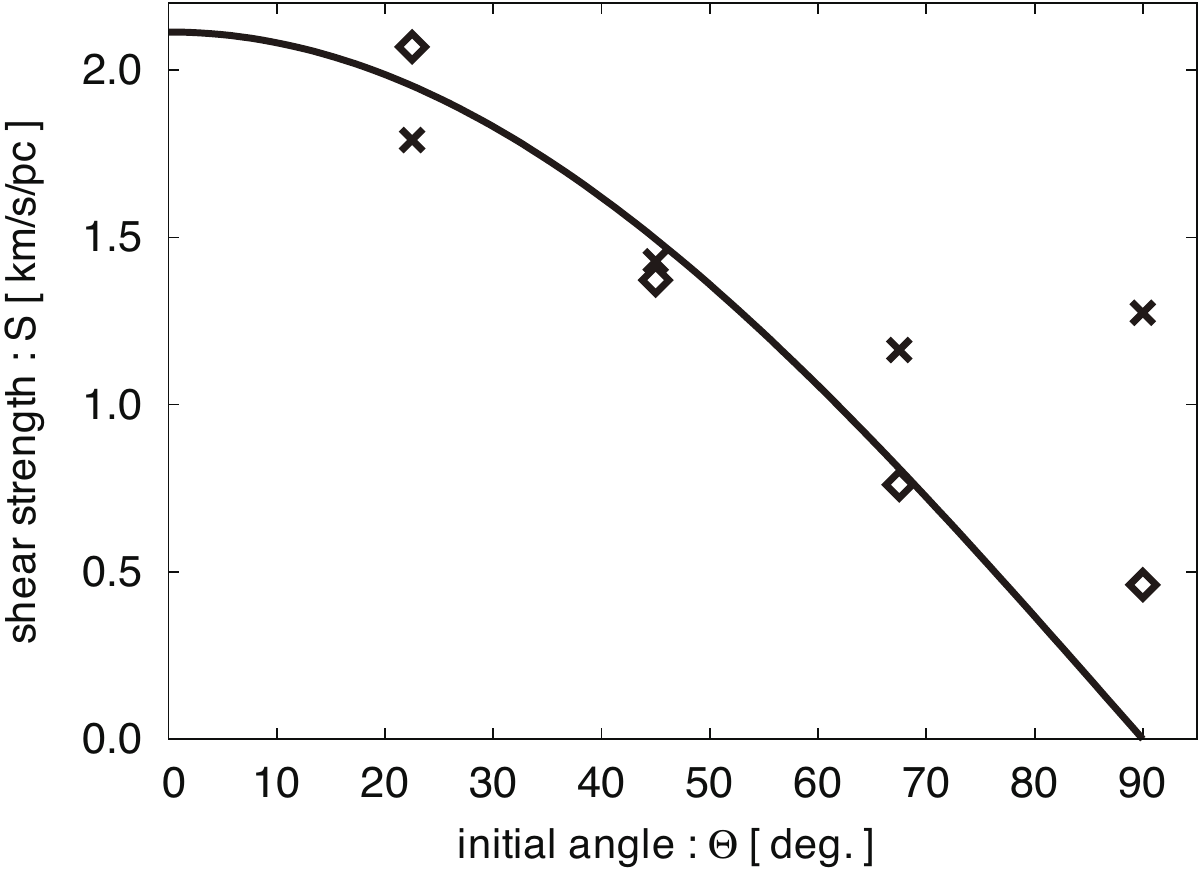}
\caption{\label{f7}
Shear strength $S$ as a function of $\Theta$.
Diamonds correspond to the models $\Theta$22V0-$\Theta$90V0 and crosses correspond to the models $\Theta$22V5-$\Theta$90V5.
Theoretical model curve eq. (\ref{eqSR}) is plotted as a solid line, where the coefficient $2.11$ km s$^{-1}$ pc$^{-1}$ prior to $\cos\Theta$ is determined by the least square fitting to the results of models $\Theta$22V0-$\Theta$90V0.
}\end{figure}

To reinforce the above discussion, we compute the velocity shear strength $S$ along the magnetic field:
\begin{equation}
 S\equiv \langle|\frac{\vec{v}\,(\vec{r}+\vec{l}_{B\perp}) - \vec{v}\,(\vec{r}-\vec{l}_{B\perp})}{|\vec{l}_{B\perp}(\vec{r})|}\,\cdot\,\frac{\vec{B}\,(\vec{r})}{|\vec{B}\,(\vec{r})|}|\rangle,
\end{equation}
where an average is taken in the region $x\in(8\mbox{ pc},12\mbox{ pc})$ and $\vec{l}_{B\perp}$ is a vector perpendicular to the local magnetic field.
To save computation time, we restrict $\vec{l}_{B\perp}$ to be in the $x$-$y$ plane and its length to be 1 pc.
This is reasonable because the initial magnetic field is set in $x$-$y$ plane.
Figure \ref{f7} shows the shear strength $S$ as a function of $\Theta$, where diamonds correspond to the models $\Theta$22V0-$\Theta$90V0 and crosses correspond to the models $\Theta$22V5-$\Theta$90V5.
We also plot the theoretical curve eq. (\ref{eqSR}) as a solid line, where the coefficient $2.11$ km s$^{-1}$ pc$^{-1}$ prior to $\cos(\Theta)$ is determined by the least square fitting to the results of models $\Theta$22V0-$\Theta$90V0.
We see that the shear strengths in the models $\Theta$22V0-$\Theta$90V0 are well fitted to eq. (\ref{eqSR}) as expected, while those of the models $\Theta$22V5-$\Theta$90V5 seem to have constant floor due to the initial turbulence.
From Figure \ref{f5} and Figure \ref{f7}, we find that the shear strength $S\sim 1$ km s$^{-1}$ pc$^{-1}$ is the threshold value for the linear structures to be aligned to the orientation of local magnetic field.
This value is quite reasonable, because the RHT with $D_W=3$ pc detects the linear structure longer than 3 pc, and the H\,{\footnotesize I} clouds of size $\sim 1$ pc can be stretched more than 3 pc long in 3 Myr (duration of simulations) under the shear flow with $S>1$ km s$^{-1}$ pc$^{-1}$.

\section{Summary and Discussion}
In this paper, we have studied the formation of the CNM in the shock-compressed layer similar to that associated with the local bubble shell, using three-dimensional MHD simulations with the effects of interstellar cooling/heating and thermal conduction.

Our findings can be summarized as follows:
A magnetized thermally unstable gas layer is created behind the shock wave in which fragments of CNM are formed via the thermal instability.
This basic evolution is essentially the same as that pointed out in our previous studies using two-dimensional simulations by Inoue \& Inutsuka (2008; 2009).
Using the RHT scheme developed by Clark et al.~(2014), we identified linear features in the simulated column density structures.
We have found that the orientation between the linear structures and local magnetic field is aligned very well when shear flows along a magnetic field ($\partial\,v_{B\parallel}/\partial l_{B\bot}$) is larger than some critical value ($\sim 0.6$ km s$^{-1}$ pc$^{-1}$) in the postshock layer.
When we reduced the shear strain along magnetic field by changing the initial direction of magnetic field, we observed that a number of linear structures lie perpendicular to the magnetic field increase.
In the cases with realistic level of initial upstream turbulence ($\Delta v_0=$5 km s$^{-1}$), we found that the CNM linear structure always tends to align with local magnetic field independent of the initial upstream orientation of the magnetic field.
The dispersion of the angular distribution is consistent with that measured by Planck collaboration XXXII (2016).

Given the above results of simulations and synthetic observations, we conclude that H\,{\footnotesize I} fibers identified by Clark et al. (2014) would be structures embedded in the H\,{\footnotesize I} shell of the local bubble as those authors proposed.
We also conclude that the formation mechanism of the linear H\,{\footnotesize I} fibers or filaments along the magnetic field would be the turbulent shear strain in agreement with the analysis by Hennebelle (2013).

The relation between the orientation of magnetic fields and structures in molecular clouds has been studied by many authors (e.g., Li et al. 2014; Malinen et al. 2016; Planck Collaboration XXXV 2016; Federrath 2016).
Planck collaboration XXXV (2016) showed that structures in molecular clouds tend to align with local magnetic field lines for low column density structures ($N_{\rm H}\lesssim 10^{21.7}$ cm$^{-2}$), while high column density structures tend to be perpendicular.
They discussed that the orientation shift would be caused by self-gravity, which piles up material along the magnetic field if it is relatively strong.
In this paper we have shown that the filamentary structures perpendicular to the magnetic field can be created by cooling contraction even in lower density atomic ISM, i.e., a magnetic-field aligned gas accumulation is possible even without self-gravity (see, the result of the model $\Theta$90V0).
In addition to this, Inoue \& Fukui (2013), Vaidya et al. (2013) and Inutsuka et al. (2015a) demonstrated that focusing flows induced behind a curved MHD fast shock create dense molecular filaments that are perpendicular to the local magnetic field.
Further theoretical and observational studies are necessary to understand the formation and evolution of the perpendicular dense filaments, since they are supposed to be the sites of star formation (Andre et al. 2014, Inutsuka et al. 2015a).

\acknowledgments
T.I. thanks  A. Lazarian, J. D. Soler, and P. Andr\'e for fruitful discussions.
The numerical computations were carried out on XC30 system at the Center for Computational Astrophysics (CfCA) of National Astronomical Observatory of Japan.
This work is supported by Grant-in-aids from the Ministry of Education, Culture, Sports, Science, and Technology (MEXT) of Japan (15K05039, 26287030).

\end{document}